\documentclass[journal]{IEEEtran}

\usepackage[usenames]{color}
\usepackage{balance}
\usepackage{amsmath}
\usepackage{amssymb}
\usepackage{multirow}
\usepackage{cite}
\usepackage{graphicx}
\usepackage{hyperref}
\usepackage{fancyhdr}

\usepackage{epsf}

\begin{document}

\title{Performance Comparisons of Geographic Routing Protocols in Mobile Ad Hoc Networks}
\author{Don Torrieri,~\IEEEmembership{Senior~Member,~IEEE,} Salvatore Talarico,~\IEEEmembership{Student Member,~IEEE,}
\and and \\ Matthew C. Valenti,~\IEEEmembership{Senior~Member,~IEEE.}
\thanks{Manuscript received Feb. 19, 2015; revised May. 30, 2015; revised Jul. 20, 2015; accepted August 28, 2015.  Date of publication XXX. XX, 2015. Date of current version XXX. XX, 2015.  The associate editor coordinating the review of this paper and approving it for publication was J. Widmer.}
\thanks{Portions of this paper were presented at the IEEE Military Communications Conference, 2013 and 2014.}
\thanks{D. ~Torrieri is with the US Army Research Laboratory,
Adelphi, MD (email: don.j.torrieri.civ@mail.mil).}
\thanks{ S.~Talarico and M.~C.~Valenti are with West Virginia University, Morgantown, WV, U.S.A. (email:salvatore.talarico81@gmail.com; valenti@ieee.org).}
\thanks{Digital Object Identifier 15.0162/TCOMM.2015.XX.XXXXXX}
}
\maketitle

\pagestyle{fancy}
\fancyhead[RO,LE]{\small\thepage}
\fancyhead[LO]{\small IEEE TRANSACTIONS ON COMMUNICATIONS, ACCEPTED FOR PUBLICATION}
\fancyhead[RE]{\small TORRIERI et al.: PERFORMANCE COMPARISONS OF GEOGRAPHIC ROUTING PROTOCOLS IN MOBILE AD HOC NETWORKS}
\fancyfoot[L,R,C]{}
\renewcommand{\headrulewidth}{0pt}

\begin{abstract}
Geographic routing protocols greatly reduce the requirements of topology
storage and provide flexibility in the accommodation of the dynamic behavior
of mobile ad hoc networks. This paper presents performance evaluations and
comparisons of two geographic routing protocols and the popular AODV protocol.
The tradeoffs among the average path reliabilities, average conditional
delays, average conditional numbers of hops, and area spectral efficiencies
and the effects of various parameters are illustrated for finite ad hoc
networks with randomly placed mobiles. This paper uses a dual method of
closed-form analysis and simple simulation that is applicable to most routing
protocols and provides a much more realistic performance evaluation than has
previously been possible. Some features included in the new analysis are
shadowing, exclusion and guard zones, distance-dependent fading, and
interference correlation.

\end{abstract}


\section{Introduction}

\PARstart{M}{obile}  ad hoc networks often use the \textit{ad-hoc on-demand distance-vector}
(AODV) routing protocol \cite{perk}, which discovers and maintains multihop
paths between source mobiles and destination mobiles. However, these paths are
susceptible to disruption due to changes in the fading, terrain, and
interference, and hence the routing overhead requirements are high. An
alternative class of routing protocols that do not maintain established routes
between mobiles are the geographic routing protocols. These protocols require
only a limited amount of topology storage by mobiles and provide flexibility
in the accommodation of the dynamic behavior of ad hoc networks (e.g.,
\cite{cad} - \cite{ghaf} and the many references therein).

Among the many varieties of geographic routing protocols, two representative
ones are evaluated in this paper: \textit{greedy forwarding}, which uses
beacons, and \textit{maximum progress routing}, which is contention based. The
tradeoffs among the average path reliabilities, average conditional delays,
average conditional numbers of hops, and area spectral efficiencies and the
effects of various parameters are illustrated for finite ad hoc networks with
randomly placed mobiles. A comparison is made with the AODV routing protocol
to gain perspective about the advantages and disadvantages of geographic routing.

There has been extensive recent research directed toward providing insights
into the tradeoffs among the reliabilities, delays, and throughputs of mobile
ad hoc networks with multihop routing (e.g., \cite{zan} - \cite{and}).
However, the mathematical models and their associated assumptions have not
been adequate for obtaining reliable results. Much of this research uses
network models based on stochastic geometry (e.g., \cite{haen} - \cite{stoy})
with the spatial distribution of the mobiles following a Poisson point
process, and simplifying but unrealistic restrictions and assumptions. One of
the principal problems associated with the models based on stochastic geometry
is that they assume an infinitely large network with an infinite number of
mobiles so that routing in the interior of the network cannot be distinguished
from routing that includes a source or destination mobile near the perimeter
of the network. The homogeneous Poisson point process does not account for the
dependencies in the placement of mobiles, such as the existence of exclusion
and guard zones \cite{tor1} that ensure a minimum spatial separation between
mobiles. Typical unrealistic restrictions are the absence of shadowing, the
neglect of thermal noise, and the identical fading conditions for each link.
Typical unrealistic assumptions are the independence of the success
probabilities of paths from the source to the destination even when paths
share the same links and the limiting of the number of end-to-end
retransmissions rather than link retransmissions. In this paper, all of these
unrealistic restrictions and assumptions are eliminated.

In this paper, rigorously derived closed-form expressions of outage
probabilities based on the methodology of \cite{tor}, which we call
\emph{deterministic geometry}, are combined with simple and rapid simulations
that allow additional network features to be considered. The simulation allows
the compilation of statistical characteristics of routing without assumptions
about the statistical independence of possible paths. During each simulation
trial, the topology is fixed, and we compute outage probabilities and
performance measures, and then we average over many topologies. Within each
topology, mobiles can be placed according to any distribution, and we focus on
uniform clustering with exclusion and guard zones. During each simulation
trial, paths for message delivery are selected by using the closed-form
expression for the per-link outage probability to determine which paths are
possible, and the delay associated with each available link is determined.
Using these paths and averaging over many topologies, the dependences of the
path reliability, area spectral efficiency, average message delay, and average
number of hops on network parameters such as the source-destination distance,
maximum number of transmission attempts per link, and density of mobiles are
evaluated. The work presented here is an extension of our preliminary work
\cite{Torrieri:2013,Torrieri:2014}. For instance, in \cite{Torrieri:2013} an
earlier form of the methodology is used to analyze three non-geographic
routing protocols. In \cite{Torrieri:2014}, preliminary results for geographic
protocols are presented, but they do not account for interference correlation.
The present paper provides a deeper analysis by considering the effects of
interference correlation, the maximum number of retransmissions, the spreading
factor, the contention density, and the relay density.

Among the features of our analysis and simulation that distinguish it from
those by other authors are the following:

1. Distinct links do not necessarily experience identically distributed
fading. For example, a distance-dependent fading model is adopted in Section IV.

2. Source-destination pairs are not assumed to be stochastically equivalent.
For example, if a source or destination is located near the perimeter of the
network, the interference and hence the routing characteristics are different
from those computed for source-destination pairs near the center of the network.

3. There are no assumptions of independent path selection or path success
probabilities. If a link fails, then all potential paths that share that link
also fail.

4. The shadowing over the link from one mobile to another can be modeled
individually, as required by the local terrain. For computational simplicity
in the example network of Section IV, the shadowing is assumed to have a
lognormal distribution.

5. The presence of thermal noise, which is integrated into the analysis, is
important when the mobile density, and hence the interference, is moderate or low.

6. The routing protocols do not depend on predetermined routes. Instead, they
use a more realistic \emph{dynamic route selection} that may include a
path-discovery phase using request packets and acknowledgements and a
message-delivery phase.

7. Due to the use of an accurate closed-form expression for the outage
probability in the presence of fading and interference, the simulation does
not need to draw random variables representing the fading and interference conditions.

8. Interference correlation \cite{Ganti:2009}, which is an artifact of the
fixed positions of the potentially interfering nodes, is naturally taken into
account because the topology is fixed for each simulation trial.

The methodology has great generality and can be applied to the performance
evaluation of most other routing protocols and types of communication networks
and environments within mobile ad hoc networks. By varying the values of
several key parameters that are defined in the paper, the essence of many
kinds of networks can be captured. Nevertheless, there are some limitations to
the approach that arise primarily from fixing the topology. For instance,
networks with very high mobility (i.e., if there is significant movement
during an end-to-end transmission) or that use store-carry-and-forward
protocols cannot be immediately accommodated. However, such networks could be
handled through an extension of the proposed methodology involving the
incorporation of a discrete-time mobility model. For ease of exposition, the
paper focuses on unicast transmission, but multicast protocols are an obvious extension.

The remainder of this paper is organized as follows. Section \ref{Sec:NetworkModel} presents the network model, featuring an equation for the outage probability for a
link between two mobiles, provides a description of the network simulator, and discusses the issue of interference correlation. Section \ref{Sec:Routing} describes three routing protocols,
the implementation of path selection, and the performance metrics used to
evaluate and compare these protocols. In Section \ref{Sec:NumericalResults}, numerical results are presented for a typical large network. Finally, the paper concludes in Section \ref{Sec:Conclusions}.

\section{Network Model and Simulator} \label{Sec:NetworkModel}

\subsection{Network Model}

The network comprises $M+2$ half-duplex mobiles in an arbitrary two- or
three-dimensional region. The variable $X_{i}$ represents both the $i^{th}$
mobile and its location, and $||X_{j}-X_{i}||$ is the distance from the
$i^{th}$ mobile to the $j^{th}$ mobile. Mobile $X_{0}$ serves as the reference
transmitter or message source, and mobile $X_{M+1}$ serves as the reference
receiver or message destination. The other $M$ mobiles $X_{1},...,X_{M}$ are
potentially relays or sources of interference. Each mobile uses a single
omnidirectional antenna.

\emph{Exclusion zones} surrounding the mobiles, which ensure a minimum
physical separation between two mobiles, have radii set equal to
$r_{\mathsf{ex}}.$ The mobiles are uniformly distributed throughout the
network area outside the exclusion zones, according to a \textit{uniform
clustering} model \cite{tor1}.

The mobiles of the network transmit asynchronous quadriphase direct-sequence
signals. For such a network, interference is reduced after despreading by the
factor $G/h(\tau_{o})$, where $G$ is the \emph{processing gain} or
\emph{spreading factor}, and $h(\tau_{o})$ is the chip factor \cite{tor},
which is a function of the chip waveform and the timing offset $\tau_{o}$ of
the interference spreading sequence relative to that of the desired or
reference signal. Since only timing offsets modulo-$T_{c}$ are relevant,
$0\leq\tau_{o}<T_{c}.$ If $\tau_{o}$ is assumed to have a uniform distribution
over [0, $T_{c})$ and the chip waveform is rectangular, then the expected
value of $h(\tau_{o})$ is 2/3. It is assumed henceforth that $G/h(\tau_{o})$
is a constant equal to $G/h$ for all mobiles in the network.

Let ${P}_{i}$ denote the received power from $X_{i}$ at the reference distance
$d_{0}$ before despreading when fading and shadowing are absent. After the
despreading, the power of $X_{i}$'s signal at the mobile $X_{j}$ is
\begin{equation}
\rho_{i,j}=\tilde{P}_{i}g_{i,j}10^{\xi_{i,j}/10}f\left(  ||X_{j}%
-X_{i}||\right)  \label{1}%
\end{equation}
where $\tilde{P}_{i}={P}_{i}$ for the desired signal, $\tilde{P}_{i}=h{P}%
_{i}/G$ for an interferer, $g_{i,j}$ is the power gain due to fading,
$\xi_{i,j}$ is a shadowing factor, and $f(\cdot)$ is a path-loss function. The
path-loss function is expressed as the power law
\begin{equation}
f\left(  d\right)  =\left(  \frac{d}{d_{0}}\right)  ^{-\alpha}\hspace
{-0.45cm},\,\,\text{ \ }d\geq d_{0} \label{2}%
\end{equation}
where $\alpha\geq2$ is the path-loss exponent, and $d_{0}$\ is a reference
distance within the near-field radius such that $r_{\mathsf{ex}}\geq d_{0}.$

The \{$g_{i,j}\}$ are independent with unit-mean but are not necessarily
identically distributed; i.e., the channels from the different $\{X_{i}\}$ to
$X_{j}$ may undergo fading with different distributions. For analytical
tractability and close agreement with measured fading statistics, Nakagami
fading is assumed, and $g_{i,j}=a_{i,j}^{2}$, where $a_{i,j}$ is Nakagami with
parameter $m_{i,j}$. It is assumed that the \{$g_{i,j}\}$ remain fixed for the
duration of a transmission but vary independently from transmission to
transmission (block fading).

In the presence of shadowing with a lognormal distribution, the $\{\xi
_{i,j}\}$ are independent zero-mean Gaussian random variables with variance
$\sigma_{s}^{2}$. For ease of exposition, it is assumed that the shadowing
variance is the same for the entire network, but the results may be easily
generalized to allow for different shadowing variances over parts of the
network. In the absence of shadowing, $\xi_{i,j}=0$. While the fading may
change from one transmission to the next, the shadowing remains fixed for the
entire session.

Each mobile may serve as either a potential relay or a potential source of interference.
The \emph{service probability} $\mu_{i}$ is defined as the probability that
mobile $X_{i}$ can serve as a relay along a path from a source to a
destination.  A Bernoulli variable with probability $\mu_{i}$ is used to
determine if $X_{i}$ is a potential relay, and if it is not, then it is a
potential interferer.
A mobile may not be able to serve as a relay in a path from
$X_{0}$ to $X_{M+1}$ because it is already receiving a transmission, is
already serving as a relay in another path, is transmitting, or is otherwise unavailable.

With \emph{interference probability} $p_{i}$, a potentially interfering
$X_{i}$ transmits in the same time interval as the desired signal. The
$\{p_{i}\}$ can be used to model the servicing of other streams, controlled
silence, or failed link transmissions and the resulting retransmission
attempts. The interference transmitted by a potential interferer is
independent from one slot to the next; hence, an Aloha medium access control
protocol is assumed. Mobiles $X_{0}$ and $X_{M+1}$ do not cause interference,
nor do the potential relays. Let $\mathcal{S}$ denote the indices of the
mobiles that actually transmit interference during a time slot. Note that the
composition of $\mathcal{S}$ is fixed for this time slot, but may vary from
one slot to the next.

Let $\mathcal{N}$ denote the noise power. Since the despreading does not
significantly affect the desired-signal power, (\ref{1}) and (\ref{2}) imply
that the instantaneous signal-to-interference-and-noise ratio (SINR) at the
mobile $X_{j}$ for a desired signal from a relay or source mobile $X_{k}$ is
\begin{equation}
\gamma_{k,j}=\frac{g_{k,j}\Omega_{k,j}}{\displaystyle\Gamma^{-1}+\sum
_{i\in\mathcal{S}}g_{i,j}\Omega_{i,j}}%
\end{equation}
where
\begin{equation}
\Omega_{i,j}=%
\begin{cases}
10^{\xi_{k,j}/10}||X_{j}-X_{k}||^{-\alpha} & i=k\\
\displaystyle\frac{h{P}_{i}}{GP_{k}}10^{\xi_{i,j}/10}||X_{j}-X_{i}||^{-\alpha}
& i\in\mathcal{S}%
\end{cases}
\end{equation}
is the normalized power of $X_{i}$ at $X_{j}$, and $\Gamma=d_{0}^{\alpha}%
P_{k}/\mathcal{N}$ is the SNR when $X_{k}$ is at unit distance from $X_{j}$
and fading and shadowing are absent.

The \emph{outage probability} quantifies the likelihood that the interference,
shadowing, fading, and noise will be too severe for useful communications.
Outage probability is defined with respect to an SINR threshold $\beta$, which
represents the minimum SINR required for reliable reception. In general, the
value of $\beta$ depends on the choice of coding and modulation. An
\emph{outage} occurs when the SINR falls below $\beta$. Let
$\boldsymbol{\Omega}_{j}=\{\Omega_{0,j},...,\Omega_{M,j}\}$ represent the set
of normalized powers at $X_{j}$. Conditioning on $\boldsymbol{\Omega}_{j}$,
the \emph{outage probability} of the link from $X_{k}$ to receiver $X_{j}$ is
\begin{equation}
\epsilon_{k,j}=P\left[  \gamma_{k,j}\leq\beta\mid\boldsymbol{\Omega}%
_{j}\right]  .
\end{equation}

The conditioning enables the calculation of the outage probability for every
link of any specific or deterministic network geometry, which cannot be done
using tools based on stochastic geometry. Restricting the Nakagami parameter
$m_{k,j}$ of the channel between the relay $X_{k}$ and receiver $X_{j}$ to be
integer-valued, the outage probability conditioned on $\boldsymbol{\Omega}%
_{j}$ is found by using deterministic geometry \cite{tor} to be
\begin{equation}
\epsilon_{k,j}=1-e^{-\beta_{k,j}z}\sum_{s=0}^{m_{k,j}-1}{\left(  \beta
_{k,j}z\right)  }^{s}\sum_{t=0}^{s}\frac{z^{-t}H_{t,j}}{(s-t)!} \label{7}%
\end{equation}
where $\beta_{k,j}=\beta m_{k,j}/\Omega_{k,j}$, $z=\Gamma^{-1},$
\begin{equation}
H_{t,j}=\mathop{ \sum_{\ell_i \geq 0}}_{\sum_{i\in\mathcal{S}}\ell_{i}=t}%
\prod_{i\in\mathcal{S}}{\mathsf{G}}_{\ell_{i}}(i,j) \label{8}%
\end{equation}
the summation in (\ref{8}) is over all sets of indices that sum to $t$,
\vspace{-0.2cm}
\begin{equation}
\mathsf{G}_{\ell}(i,j)=%
\begin{cases}
\Psi_{i,j}^{m_{i,j}} & \ell=0\\
\frac{\Gamma(\ell+m_{i,j})}{\ell!\Gamma(m_{i,j})}\left(  \frac{\Omega_{i,j}%
}{m_{i,j}}\right)  ^{\ell}\Psi_{i,j}^{m_{i,j}+\ell} & \ell>0
\end{cases}
\end{equation}
and
\begin{equation}
\Psi_{i,j}=\left(  \beta_{k,j}\frac{\Omega_{i,j}}{m_{i,j}}+1\right)
^{-1},\text{ \ }i\in\mathcal{S}. \label{10}%
\end{equation}

\subsection{Network Simulation} \label{subsec:NetworkSimulator}

The simulator is organized into four layers, each of which emulates a
particular random feature of the network. The layers are implemented as nested
\emph{for} loops. The top layer handles the random network topology (i.e., the
location of all mobiles in the network), the second layer is concerned with
classifying each mobile as either a potential source of interference or a
potential relay, the third layer is concerned with determining which of the
potential interferers actually transmits (assuming an Aloha protocol with
random access probability $p_{i}$ is used), while the bottom layer determines
which links are actually in an outage. The routing is also handled at the
bottom layer.

During each iteration of the top layer, a network topology is generated by
placing the locations of the mobiles according to the deterministic geometry
model. First the source and destination mobiles are placed in fixed positions,
and then one by one, the location of each of the $M$ remaining mobiles is
drawn according to a uniform distribution within the network region. However,
if an $X_{i}$ falls within the exclusion zone of a previously placed mobile,
then it has a new random location assigned to it as many times as necessary
until it falls outside all exclusion zones. This loop is run $\Upsilon$ times,
once per topology.

During each iteration of the second layer, which is run $K_{t1}$ times, each
mobile is independently marked as either being a potential relay (with
probability $\mu_{i}$) or a potential source of interference. Those mobiles
that are marked as potential relays cannot transmit interference, and are
forced to have transmission probability $p_{i}=0$.

During each iteration of the third loop, which is run $K_{t2}$ times, the set
of transmitting interferers $\mathcal{S}$ is found for each time slot, up to
some maximum time. Each $\mathcal{S}$ is found by marking each potential
interferer as transmitting by drawing a Bernoulli variable with probability
$p_{i}$. Then, for each time slot $t$, a matrix is found containing the outage
probabilities between all mobiles that may participate in the route (i.e., all
mobiles except for the potential interferers).

During each iteration of the bottom layer, which is run $K_{t3}$ times, each
link is marked as either being in an outage or not for each time slot by
drawing a Bernoulli variable with probability equal to the corresponding entry
in the channel outage matrices. Once the links are marked as being in an
outage or not, the set of links that may be used to support a route is
identified, and the corresponding routing protocols (described in the next
section) may be implemented. Counters are updated to keep track of the key
network performance metrics.

\subsection{Interference Correlation}

For a given network topology and service model, there is a common set of
mobiles that may produce interference. These mobiles are in locations that are
relatively fixed when compared to the timescales of communications. Because of
the common randomness in the locations of potentially interfering mobiles,
there is correlation in the interference. The correlation is both temporal,
because subsequent interfering transmissions come from subsets of the same
common set of mobiles, and spatial, due to the common locations of interfering
mobiles within a given time slot. The spatio-temporal correlation exists even
when the mobiles use Aloha as the MAC protocol, which produces
\emph{transmissions} that are locally uncorrelated. While interference
correlation is usually neglected, it has become a subject of recent interest
\cite{Ganti:2009}.

The network model and simulator presented here naturally accounts for
interference correlation. This is because during each iteration of the top
layer of the simulation, the topology is fixed. Hence, the interfering mobiles
selected at the third layer of the simulation is constrained to be drawn from
this set. It follows that the calculation of outage probability properly
accounts for interference correlation.

Here, the simulator assumes a simple Aloha protocol is used by each potential
interferer. We note that other mobiles in the network are likely to be
participating in some other, unknown, path. The protocol may be the same or
different than the protocol used by the reference path. While it is possible
to extend the methodology herein to simulate the separate routing action of
neighboring nodes, such an approach is contrary to the simple simulation
advocated by this paper, and requires assumptions to be made about the
behavior of the other nodes. In many ad hoc environments, the other nodes may
be using completely different routing protocols, may be engaged in direct
device-to-device communications, or may be sources of intentional jamming.
Rather than trying to simulate the action of each potential interferer, we
capture the behavior by appropriately setting the value of the interference
probability $p_{i}$. In particular, our investigations have found that for a
given routing protocol used by the potential interferers, there will be an
equivalent value of $p_{i}$ under our model that results in the same performance.

\section{Routing Models}\label{Sec:Routing}

\subsection{Routing Protocols}

The three routing protocols that are considered are reactive or on-demand
protocols that only seek paths from the source to the destination when needed
and do not require mobiles to store details about large portions of the network.

The AODV protocol uses an on-demand approach for finding a route during
its\emph{\ path-discovery phase,} which relies on flooding to seek the
\emph{fewest-hops path,} which is the path with the smallest number of links
or hops. The flooding diffuses request packets simultaneously over multiple
routes for the purpose of discovering a successful path to the destination
despite link failures along some potential paths. When the first request
packet reaches the destination, an acknowledgement is sent back to the source
using the fewest-hops path discovered, and each subsequent reception of
request packets is ignored. By using this path, the source sends subsequent
message packets to the destination during a \emph{message-delivery phase}.
This protocol has a very high overhead cost in establishing the fewest-hops
path during the path-discovery phase \cite{haq}, and furthermore the
fewest-hops path must be used for message delivery before changes in the
channel conditions cause an outage of one or more of its links.

Geographic protocols limit information-sharing costs by minimizing the
reliance of mobiles on topology information \cite{cad} - \cite{ghaf}. Since
geographic routing protocols make routing decisions on a hop-by-hop basis,
they do not require a flooding process for path discovery. Two geographic
routing protocols are examined: the \emph{greedy forwarding protocol} and the
\emph{maximum progress protocol. }Both geographic routing protocols assume
that each potential relay knows its physical location and the direction towards the destination.

The greedy forwarding protocol relies on \emph{beacons}, which are periodic
brief messages exchanged among mobiles that serve to identify neighboring
active mobiles and their locations. Under the greedy forwarding protocol, a
source forwards a packet to a relay that is selected from a set of
neighboring active mobiles that lie within a \emph{transmission range} of
radius $r_{\mathsf{t}}.$ The next link in the path from source $X_{0}$ to
destination $X_{M+1}$ is the link to the relay within the transmission range
that shortens the remaining distance to $X_{M+1}$ the most. There is no
path-discovery phase because the relays have the geographic information
necessary to route the messages to the destination.

The principal problems with beacons are that they generate additional
interference that degrades data packets, and the location-information may
become outdated. With \emph{beaconless routing}, these problems are
substantially reduced. Mobiles broadcast short request messages during a
path-discovery phase only when they are ready to transmit data packets. The
responses to the request messages reveal the identity and location of a
neighboring active mobile that is suitable as the next relay during the
message-delivery phase.

The maximum progress protocol is a \emph{contention-based} beaconless protocol
that comprises alternating \emph{path-discovery phases} and
\emph{message-delivery phases}. During a path-discovery phase, a single link
to a single relay is discovered. During the following message-delivery phase,
a packet is sent to that relay, and then the alternating phases resume until
the destination is reached. In a path-discovery phase, the next relay in a
path to the destination is dynamically selected at each hop of each packet and
depends on the local configuration of available relays. A source or relay
broadcasts a \emph{Request-to-Send} (RTS) message to neighboring active mobiles
that potentially might serve as the next relay along the path to the
destination. The RTS message includes the location of the transmitter. Upon
receiving the RTS, a neighboring mobile initiates a timer that has an
expiration time proportional to the remaining distance to the destination.
When the timer reaches its expiration time, the mobile sends a
\emph{Clear-to-Send} (CTS) message as an acknowledgement packet to the
transmitter. The earliest arriving CTS message causes the source or previous
relay to launch the message-delivery phase by sending the information-bearing message to the
mobile that sent that CTS message, and all other candidate mobiles receiving
that CTS message cease operation of their timers. The primary advantage of the
RTS and CTS messages is that they can be used to establish guard zones
surrounding the transmitter and receiver \cite{tor1}. Potentially interfering
mobiles that receive one of these messages within a guard zone are silenced
during the acknowledgement and message-delivery phases of the maximum progress
protocol. In contrast, while beacons enable direct message transmissions
without preliminary RTS/CTS phases, the unsilenced interference may cause
outages, thereby significantly degrading the path reliability.

\subsection{Implementation of Path Selection}

\begin{figure}[t!]
\centering
\includegraphics[width=9cm]{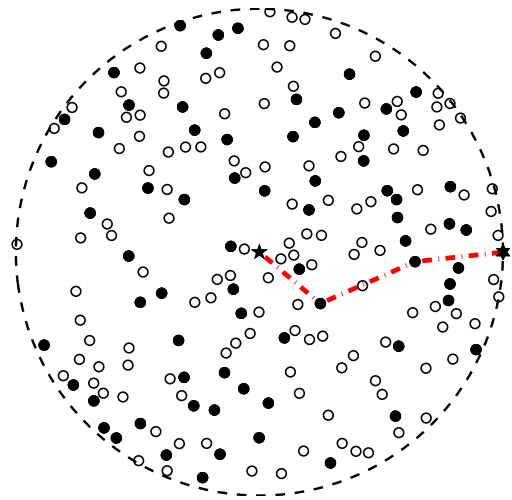}
\vspace{-0.5cm}
\caption{Typical network topology for two communicating mobiles and $M=200$
other mobiles, each of which is represented by a dot. The star at the center
of the circle represents the source $X_{0}$, and the star on the perimeter represents the
destination $X_{M+1}.$ If a dot is filled, then it may serve as a potential relay. A typical
least-hop path is indicated by the dashed line.} \label{Fig1_Topology}
\vspace{-0.25cm}
\end{figure}

A typical network topology of a large network is illustrated in Fig. \ref{Fig1_Topology}. Each
dot represents one of $M=200$ mobiles, the five-pointed star at the center of the circle
represents the source $X_{0}$, and the six-pointed star on the perimeter represents the destination
$X_{M+1}.$ If a dot is filled, then it may serve as a potential relay. A typical fewest-hops path
from $X_{0}$ to $X_{M+1}$ is indicated by the dashed line.%

A \emph{candidate link} is a link that does not experience an outage during
the path-discovery phase. For AODV, the \emph{candidate paths} from $X_{0}$ to
$X_{M+1}$ are paths that can be formed by using candidate links. The candidate
path with the fewest hops from $X_{0}$ to $X_{M+1}$ is selected as the
\emph{fewest-hops path}. This path is determined by using the \emph{Djikstra
algorithm} \cite{bru} with the unit cost of each candidate link. If two or
more candidate paths have the fewest hops, the fewest-hops path is randomly
selected from among them. If there is no set of candidate links that allow a
path from $X_{0}$ to $X_{M+1},$ then a \emph{routing failure} occurs and is
recorded. If a fewest-hops path exists, then a Monte Carlo simulation is used
to determine whether the acknowledgement packet traversing the path in the
reverse direction is successful. If it is not or if the message delivery over
the fewest-hops path fails, then a routing failure is recorded.

The geographic routing protocols have knowledge of the direction towards the
destination, and a \emph{distance criterion} is used to exclude a link from
mobile $X_{i}$ to mobile $X_{j}$ as a link in one of the possible paths from
$X_{0}$ to $X_{M+1}$ if $||X_{j}-X_{M+1}||>||X_{i}-X_{M+1}||$. These
exclusions ensure that each possible path has links that always reduce the
remaining distance to the destination. All links connected to mobiles that
cannot serve as relays are excluded as links in possible paths from $X_{0}$ to
$X_{M+1}.$ Links that have not been excluded are called \emph{eligible} links.

For the greedy-forwarding protocol there is no path-discovery phase, and the
eligible links are used to determine the greedy-forwarding path from $X_{0}$
to $X_{M+1}$ during its message-delivery phase. If no path from $X_{0}$ to
$X_{M+1}$ can be found or if the message delivery fails, a routing failure is recorded.

A \emph{two-way candidate link} is an eligible link that does not experience
an outage in either the forward or the reverse direction during the
path-discovery phase. A Monte Carlo simulation is used to determine the
two-way candidate links. For the maximum progress protocol, the two-way
candidate link starting with source $X_{0}$ with a terminating relay that
minimizes the remaining distance to destination $X_{M+1}$ is selected as the
first link in the maximum-progress path. The link among the two-way candidate
links that minimizes the remaining distance and is connected to the relay at
the end of the previously selected link is added successively until the
destination $X_{M+1}$ is reached and hence the maximum-progress path has been
determined. After each relay is selected, a message packet is sent in the
forward direction to the selected relay. If no maximum-progress path from
$X_{0}$ to $X_{M+1}$ can be found or if a message delivery fails, a routing
failure is recorded.

Each RTS or CTS message transmitted by the maximum progress protocol during
its path-discovery phase establishes a guard zone. If a candidate link exists
from the source of an RTS or CTS message to a potentially interfering mobile
within the source's guard zone, then that mobile is silenced during the
acknowledgement and message-delivery phases. Silencing of a mobile is modeled
by removing it from the set $\mathcal{S}.$ Since the maximum progress protocol
is a geographic protocol, potentially interfering mobiles know how far they
are from the source of an RTS or CTS message. If they are beyond the guard
zone, then they can ignore the message and continue their own transmissions.

In mobile ad hoc networks, the fading processes affecting different links of
the same path are not significantly correlated for two reasons. The exclusion
zones surrounding the two receivers ensure a significant physical separation.
As a result, the two receivers have much different multipath environments and
hence experience largely uncorrelated fading. The second reason is that the
time between transmissions over successive links in a path usually exceeds the
channel coherence time. These two factors decorrelate the fading over
different links of the same path.

\subsection{Performance Metrics}

Let $B$ denote the maximum number of transmission attempts over a link of the
path. During the path-discovery phases, $B=1$. During the message-delivery
phases, $B\geq1$ because message retransmissions over an established link are
feasible. For each eligible or candidate link $l=\left(  i,j\right)  $, a
Bernoulli random variable with failure probability $\epsilon_{l}$ is
repeatedly drawn until there are either $B$ failures or success after $N_{l}$
transmission attempts, where $N_{l}\leq B$. The \emph{delay of link }$l$ of
the selected path is $N_{l}T+(N_{l}-1)T_{e},$ where $T$ is the \emph{delay of
a transmission over a link}, and $T_{e}$ is the \emph{excess delay} caused by
a retransmission.

Each network topology $t$ is used in $K_{t}$ simulation trials. The \emph{path
delay} $T_{s,t}$ of a path from $X_{0}$ to $X_{M+1}$ for network topology $t$
and simulation trial $s$ is the sum of the link delays in the path during the
message-delivery phase:
\begin{equation}
T_{s,t}=\sum\limits_{l\in\mathcal{L}_{s,t}}[N_{l}T+(N_{l}-1)T_{e}].
\end{equation}
where $\mathcal{L}_{s,t}$ is the set of links constituting the path. If there
are $B$ transmission failures for any link of the selected path, then a
routing failure occurs.

If there are $F_{t}$ routing failures for topology $t$ and $K_{t}$ simulation
trials, then the \emph{probability of end-to-end success} or \emph{path}
\emph{reliability} within topology $t$ is
\begin{equation}
R_{t}=1-\frac{F_{t}}{K_{t}}. \label{13}%
\end{equation}
Let $\mathcal{T}_{t}$ denote the set of $K_{t}-F_{t}$ trials with no routing
failures. If the selected path for trial $s$ has $h_{s,t}$ links or hops, then
among the set $\mathcal{T}_{t}$, the average conditional\emph{ number of hops}
from $X_{0}$ to $X_{M+1}$ is%
\begin{equation}
H_{t}=\frac{1}{K_{t}-F_{t}}\sum\limits_{s\in\mathcal{T}_{t}}h_{s,t}.
\end{equation}
Let $T_{d}$ denote the link delay of packets during the path-discovery phase.
The average conditional\emph{ delay} from $X_{0}$ to $X_{M+1}$ during the
combined path-discovery and message-delivery phases is%
\begin{equation}
D_{t}=\frac{1}{K_{t}-F_{t}}\sum\limits_{s\in\mathcal{T}_{t}}\left(
T_{s,t}+2ch_{s,t}T_{d}\right)
\end{equation}
where $c=0$ for the greedy forwarding protocol, and $c=1$ for the maximum
progress and AODV protocols. Let $A$ denote the network area and
$\lambda=(M+1)/A$ denote the density of the possible transmitters in the
network. We define the \emph{normalized} \emph{area spectral efficiency} for
the $K_{t}$ trials of topology $t$ as
\begin{equation}
\mathcal{A}_{t}=\frac{\lambda}{K_{t}}\sum\limits_{s\in\mathcal{T}_{t}}\frac
{1}{T_{s,t}+2ch_{s,t}T_{d}}%
\end{equation}
where the normalization is with respect to the number of message bits
transmitted over a successful path. The normalized area spectral efficiency is
a measure of the maximum end-to-end throughput in the network. After computing
$R_{t},$ $D_{t},$ $H_{t},$ and $\mathcal{A}_{t}$ for $\Upsilon$ network
topologies, we can average over the topologies to compute the
\emph{topological averages: }$\overline{R,}$\ $\overline{D},$ $\overline{H},$
and $\overline{\mathcal{A}}.$

The average values of the service probability and the interference probability
are defined as
\begin{eqnarray}
\overline{\mu}
& = &
\frac{1}{M}\sum\limits_{i=1}^{M}\mu_{i} \nonumber \\
\overline{p}
& = &
\frac{1}{M}\sum\limits_{i=1}^{M}p_{i}
\end{eqnarray}
respectively. The \emph{relay density} $\lambda
\overline{\mu}$ is a measure of the average number of available relays per unit area.
The \emph{contention density\ }  $\mathbb E [ \lambda \overline{p} ]$ is a
measure of the expected number of interfering transmissions per unit area, where the expectation is with respect to the network
geometry. When the service probabilities are all the same (i.e., $\mu_i = \mu$ for all $i$), and the interference probabilities of the potential interferers are all the same (i.e., $p_i = p$ when $X_i$ is not serving as a potential relay), then the contention density is $\mathbb E [ \lambda \overline{p} ]
=\lambda p (1-\mu)$.

\section{Numerical Results}\label{Sec:NumericalResults}

A host of network topologies and parameter values can be evaluated by the
method described in Section II. Here, we consider a representative example
that illustrates the tradeoffs among the routing protocols. We consider a
network occupying a circular region with normalized radius $r_{net}=1.$ The
source mobile is placed at the origin, and the destination mobile is placed at
distance $||X_{M+1}-X_{0}||$ from it. Times are normalized by setting $T=1$.
Each transmitted power ${P}_{i}$ is equal. There are no retransmissions during
the path-discovery phases, whereas no more than $B$ retransmissions are
allowed during the message-delivery phases$.$ A \emph{distance-dependent
fading} model is assumed, where a signal originating at mobile $X_{i}$ arrives
at mobile $X_{j}$ with a Nakagami fading parameter $m_{i,j}$ that depends on
the distance between the mobiles. We set
\begin{equation}
m_{i,j}=%
\begin{cases}
3 & \mbox{ if }\;||X_{j}-X_{i}||\leq r_{\mathsf{f}}/2\\
2 & \mbox{ if }\;r_{\mathsf{f}}/2<||X_{j}-X_{i}||\leq r_{\mathsf{f}}\\
1 & \mbox{ if }\;||X_{j}-X_{i}||>r_{\mathsf{f}}%
\end{cases}
\end{equation}
where $r_{\mathsf{f}}$ is\ the \emph{line-of-sight radius}. The
distance-dependent-fading model characterizes the typical situation in which
nearby mobiles most likely are in each other's line-of-sight, while mobiles
farther away from each other are not. The severity of the fading decreases,
and hence the Nakagami parameter increases, with decreasing distance. The
transmission range $r_{t}$ defined by the greedy forwarding protocol usually
approximates or exceeds $r_{\mathsf{f}}$.

The exclusion-zone and guard-zone radii are $r_{ex}=0.05$ and $r_{\mathsf{g}%
}=0.15,$ respectively. Other fixed parameter values are $T_{e}=1.2,$
$T_{d}=0.1,$ $M=200$, $\lambda=201/\pi,$ $\beta=0$ dB, $K_{t}=K_{t1}%
K_{t2}K_{t3}=10^{6}$ $(K_{t1}=K_{t2}=K_{t3}=100),$ $\Gamma=0$ dB,
$\alpha=3.5,$ and $\Upsilon=2000.$ The service probabilities are $\mu_{i}%
=\mu,$ whereas the interference probabilities are $p_{i}=p$ when $p_{i}\neq0.$
Therefore, $\lambda\mu$ is the relay density, and $\mathbb E [ \lambda \overline{p} ] = \lambda p (1-\mu)$ is  the
contention density. Unless otherwise stated, $G/h=96,\alpha=3.5,$ $B=4$,
$r_{\mathsf{f}}=0.2,$ $\mu=0.4,$ and $p=0.3$. When shadowing is present, it
has a lognormal distribution with $\sigma_{s}=8$ dB. However, the transmitted
packets encounter the same shadowing in both directions over the same link
during both routing phases. %

\begin{figure}[t!]%
\centering
\includegraphics[width=9cm]{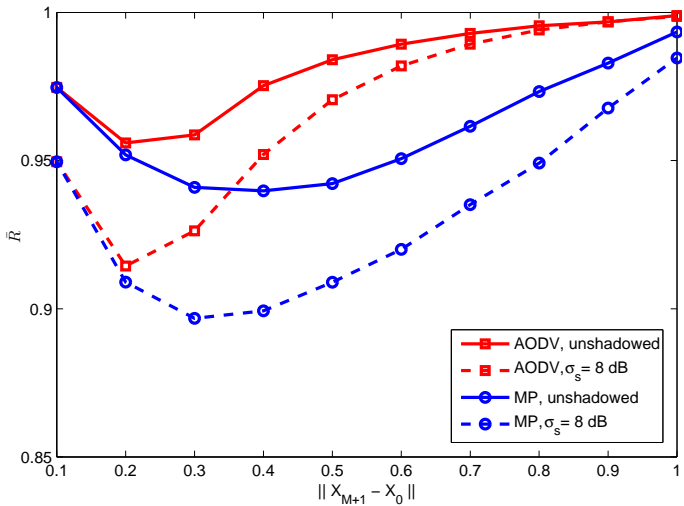}
\vspace{-0.5cm}
\caption{Average path reliability for request packets of AODV and MP protocols
as a function of the distance between source and destination. \label{Fig2_Reliability1}}
\end{figure}

\begin{figure}[t!]%
\centering
\includegraphics[width=9cm]{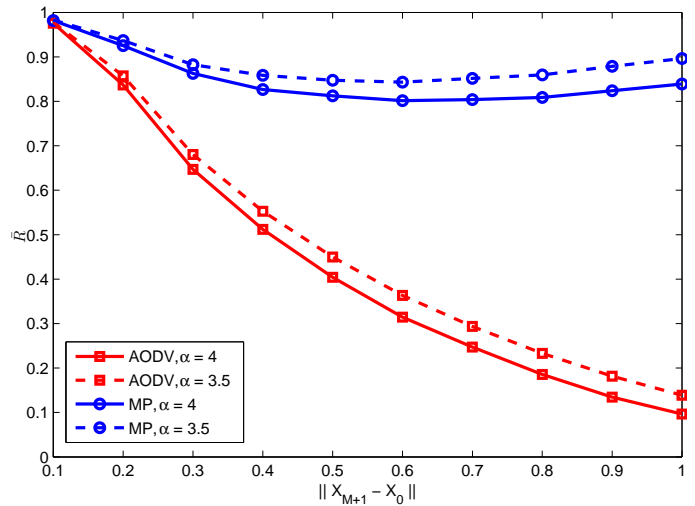}
\vspace{-0.5cm}
\caption{Average path reliability for acknowledgements of AODV and MP
protocols as a function of the distance between source and destination. \label{Fig3_Reliability2}}
\end{figure}

Fig. \ref{Fig2_Reliability1} and Fig. \ref{Fig3_Reliability2}  display the average path reliabilities of the request packets
and acknowledgement packets, respectively, for the complete selected paths
during the path-discovery phases of the AODV and maximum progress (MP)
protocols. Fig. \ref{Fig2_Reliability1} depicts the reliabilities both with and without shadowing
as a function of the source-destination distance $||X_{M+1}-X_{0}||.$
Shadowing is assumed in Fig. \ref{Fig3_Reliability2} and all subsequent figures. Fig. \ref{Fig2_Reliability1} shows an
initial decrease and then an increase in average path reliability as the
source-destination distance increases. This variation occurs because at short
distances, there are very few relays that provide forward progress, and often
the only candidate or two-way candidate link is the direct link from source to
destination. As the distance increases, there are more candidate and two-way
candidate links, and hence the network benefits from the diversity.
Furthermore, as the destination approaches the edge of the network, the path
discovery benefits from a decrease in interference at the relays that are
close to the destination. Fig.  \ref{Fig2_Reliability1} shows that during the request stage, the
AODV protocol provides the better path reliability because it constructs
several partial paths before the complete path is determined.

Since the relays are already determined in Fig. \ref{Fig3_Reliability2}, the maximum progress
protocol shows only a mild improvement with increasing source-destination
distance, and this can be attributed almost entirely to the edge effect. It is
observed in Fig. \ref{Fig3_Reliability2} that the AODV protocol has a relatively poor path
reliability during the acknowledgement stage, which is due to the fact that a
specified complete path must be traversed in the reverse direction, where the
interference and fading may be much more severe. The maximum progress protocol
does not encounter the same problem because the links in its paths are
selected one-by-one with the elimination of links that do not provide
acknowledgements. Although both the shadowing and the path-loss exponent
$\alpha$ affect both the packets and the interference signals, the two figures
indicate that the overall impact of more severe propagation conditions is
detrimental for all distances.

\begin{figure}[t!]%
\centering
\includegraphics[width=9cm]{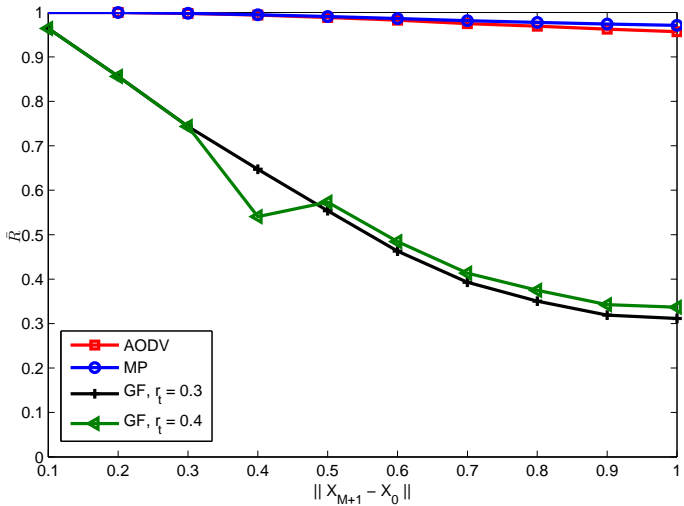}
\vspace{-0.5cm}
\caption{Average path reliability for message-delivery phase of each routing
protocol as a function of the distance between source and destination. \label{Fig4_Reliability3}}
\end{figure}

Fig. \ref{Fig4_Reliability3} displays the average path reliabilities for the message-delivery
phases of the three protocols, assuming that the path-discovery phase, if
used, has been successful. The figure illustrates the penalties incurred by
the greedy forwarding (GF) protocol because of the absence of a path-discovery
phase that eliminates links with excessive shadowing, interference, or fading
and creates guard zones for the message-delivery phase. The figure illustrates
the role of the transmission range $r_{t}$ in determining average path
reliability for greedy forwarding protocols. As $r_{t}$ increases, the links
in the complete path are longer and hence less reliable. However, this
disadvantage is counterbalanced by the increased number of potential relays
and the reduction in the average number of links in a complete path. When
$||X_{M+1}-X_{0}||$ increases slightly above $r_{t},$ there is a sudden jump
in the average reliability because the greedy forwarding protocol no longer
allows an attempt by the source to directly communicate with the destination
in one hop. Instead, two hops over more reliable links must be used.

\begin{figure}[t!]%
\centering
\includegraphics[width=9cm]{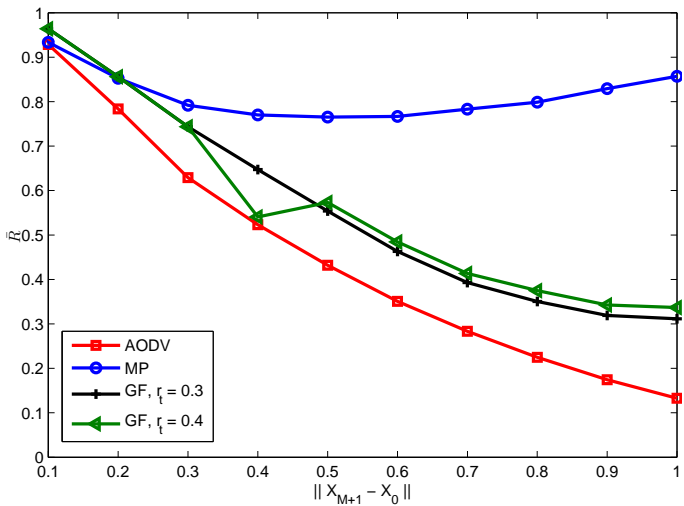}
\vspace{-0.5cm}
\caption{Average path reliability for both phases of each routing protocol as
a function of the distance between source and destination. \label{Fig5_ReliabilityTot}}
\end{figure}

Fig. \ref{Fig5_ReliabilityTot} shows the overall average path reliabilities for the combined
path-discovery and message-delivery phases of all three routing protocols. The
AODV protocol is the least reliable. The maximum progress protocol is much
more reliable than the greedy forwarding protocol if $||X_{M+1}-X_{0}||$ is
large, but is not as reliable if $||X_{M+1}-X_{0}||\leq0.2$ because of the
relatively low reliability of its request packets.

\begin{figure}[t!]%
\centering
\includegraphics[width=9cm]{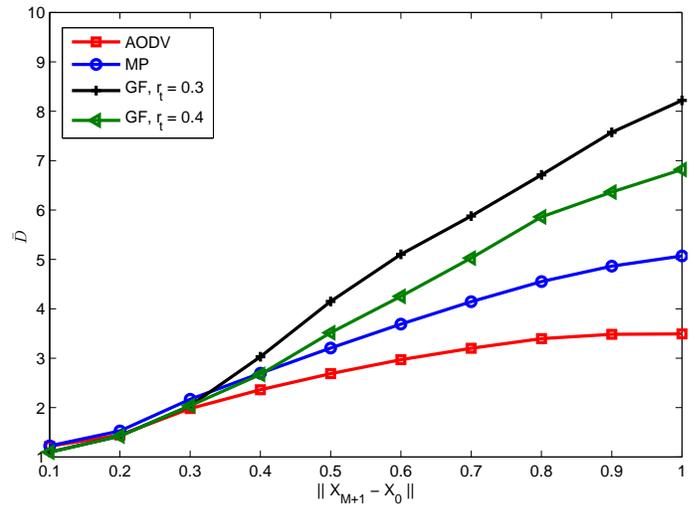}
\vspace{-0.5 cm}
\caption{Average conditional delay of each routing protocol as a
function of the distance between source and destination.}
\label{Fig6_DelayNumTra}
\end{figure}

\begin{figure}[t!]%
\centering
\includegraphics[width=9cm]{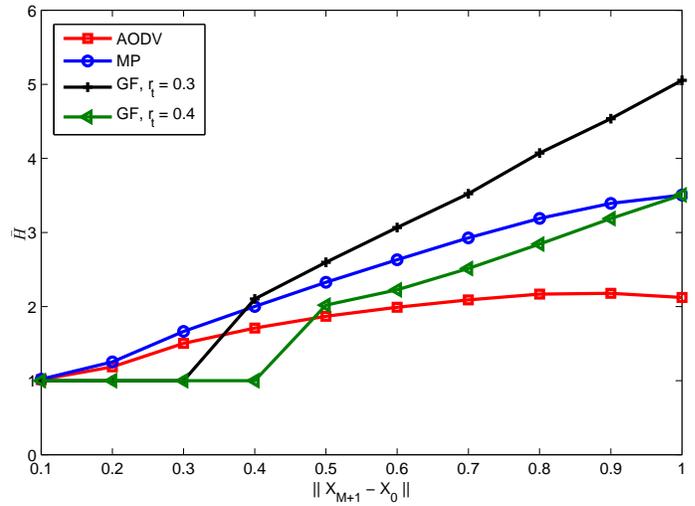}
\vspace{-0.5 cm}
\caption{Average conditional number of hops of each routing protocol as a
function of the distance between source and destination. \label{Fig7_HopsDis}}
\end{figure}

\begin{figure}[t!]%
\centering
\includegraphics[width=9cm]{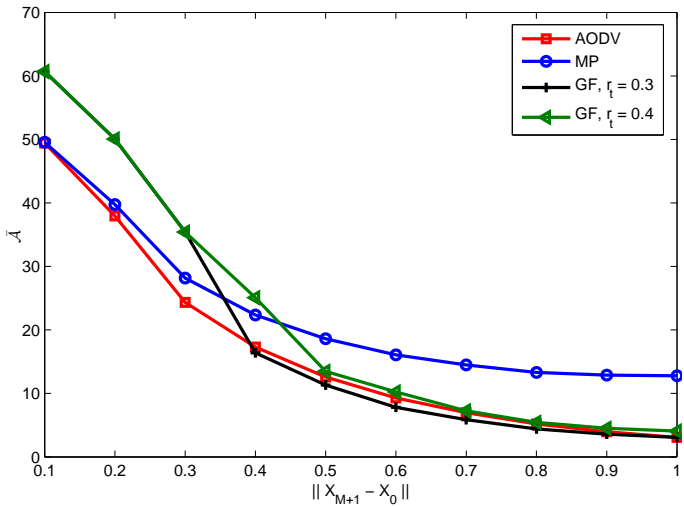}
\vspace{-0.5 cm}
\caption{Area spectral efficiency of each routing protocol as a function of
the distance between source and destination. \label{Fig8_ASPdis}}
\vspace{0.3 cm}
\end{figure}

The average conditional delay $\overline{D},$ the average conditional number
of hops $\overline{H,}$ and the normalized area spectral efficiency
$\overline{\mathcal{A}}$ for each routing protocol as a function of
$||X_{M+1}-X_{0}||$ are displayed in Fig. \ref{Fig6_DelayNumTra}
, Fig. \ref{Fig7_HopsDis}, and Fig. \ref{Fig8_ASPdis}, respectively. The
AODV protocol has the smallest $\overline{D},$ and if $||X_{M+1}-X_{0}%
||\geq0.5,$ the AODV protocol has the smallest $\overline{H}.$ The greedy
forwarding protocol has the highest $\overline{\mathcal{A}}$ if $||X_{M+1}%
-X_{0}||$ is small, whereas the maximum progress\ protocol has the highest
$\overline{\mathcal{A}}$ if $||X_{M+1}-X_{0}||$ is large. The reason is that
the greedy forwarding protocol needs only one hop for a message to reach the
destination when $||X_{M+1}-X_{0}||$ is small, whereas this protocol has
reduced reliability and significantly increased average conditional delay when
$||X_{M+1}-X_{0}||$ is large.%

\begin{figure}[t!]%
\centering
\includegraphics[width=9cm]{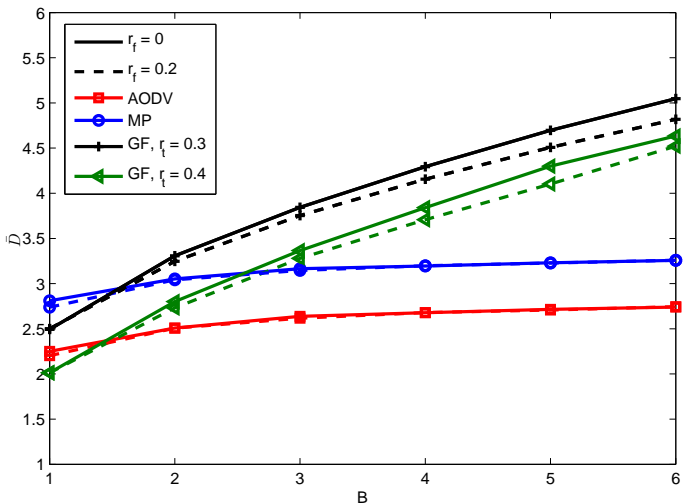}
\vspace{-0.5 cm}
\caption{Average conditional delay of each routing protocol as a
function of the maximum number of retransmissions during the message-delivery
phase when $||X_{M+1}-X_{0}||=0.5.$ \label{Fig9_DelayB}}
\end{figure}
\begin{figure}[t!]%
\centering
\includegraphics[width=9cm]{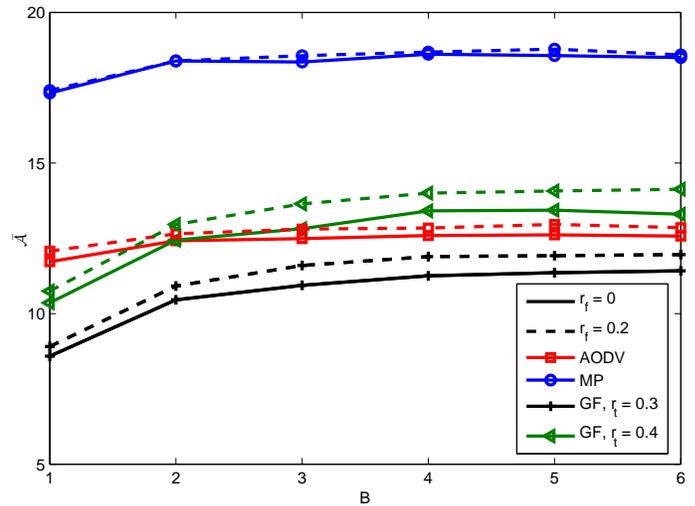}
\vspace{-0.5 cm}
\caption{Normalized area spectral efficiency of each routing protocol as a
function of the maximum number of retransmissions during the message-delivery
phase when $||X_{M+1}-X_{0}||=0.5.$ \label{Fig10_ASEB}}
\end{figure}

The average conditional delay $\overline{D}$ and the normalized area spectral
efficiency $\overline{\mathcal{A}}$ for each routing protocol as a function of
the maximum number of retransmissions during the message-delivery phase when
$||X_{M+1}-X_{0}||=0.5$ are displayed in Fig. \ref{Fig9_DelayB} and Fig. \ref{Fig10_ASEB}, respectively. The
greedy forwarding protocol has a monotonically increasing $\overline{D}$ and
path reliability as $B$ increases because paths from the source to the
destination with longer delays become viable. However, an increase in $B$ has
little effect on $\overline{D}$ for the AODV and maximum progress protocols.
An increase from $B=1$ to $B=2$ produces an increase in $\overline
{\mathcal{A}}$ for all three protocols, but further increases in $B$ produce
only a minor improvement in $\overline{\mathcal{A}}$ . The reason is that the
higher path reliabilities are offset by the increased numbers of successful
paths with longer delays. Fig. \ref{Fig9_DelayB} and Fig. \ref{Fig10_ASEB} illustrate the losses incurred by
the protocols when the fading is the more severe Rayleigh fading
($r_{\mathsf{f}}=0)$ instead of mixed fading with $r_{\mathsf{f}}=0.2.$ In all
other figures, $r_{\mathsf{f}}=0.2.$

\begin{figure}[t!]%
\centering
\includegraphics[width=9cm]{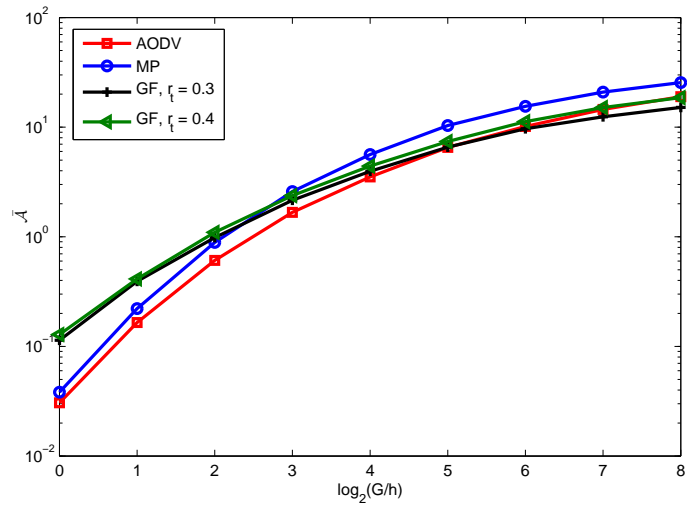}
\vspace{-0.5 cm}
\caption{Area spectral efficiency of each routing protocol as a function of
the spreading factor when $||X_{M\text{+1}}-X_{0}||=0.5$.}
\label{Fig11_ASE_G}
\vspace{0.2cm}
\end{figure}

The critical role of the spreading factor $G$ in suppressing interference is
illustrated in Fig. \ref{Fig11_ASE_G} for $||X_{M\text{+1}}-X_{0}||=0.5$. The normalized area
spectral efficiency $\overline{\mathcal{A}}$ of each protocol increases
monotonically with $G/h,$ but the rate of increase is greatest for the maximum
progress protocol$.$ When the distance is $||X_{M\text{+1}}-X_{0}||=0.5,$ the
maximum progress protocol provides the largest $\overline{\mathcal{A}}$ if
$G/h\geq8.$ In the absence of spreading, the greedy forwarding protocol
provides the largest $\overline{\mathcal{A}}.$

\begin{figure}[t!]%
\centering
\includegraphics[width=9cm]{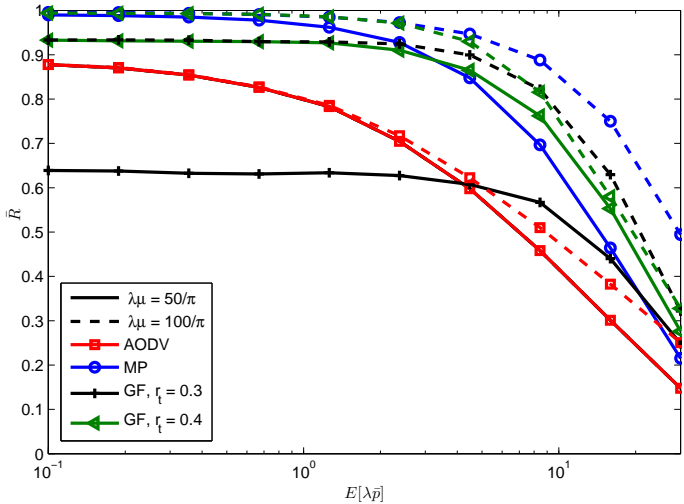}
\vspace{-0.5 cm}
\caption{Average path reliability for each routing protocol when
$||X_{M\text{+1}}-X_{0}||=0.5$ as a function of the contention
density\emph{\ }with the relay density as a parameter. \label{Fig12_Reliability_lambda}}
\end{figure}

\begin{figure}[t!]%
\centering
\includegraphics[width=9cm]{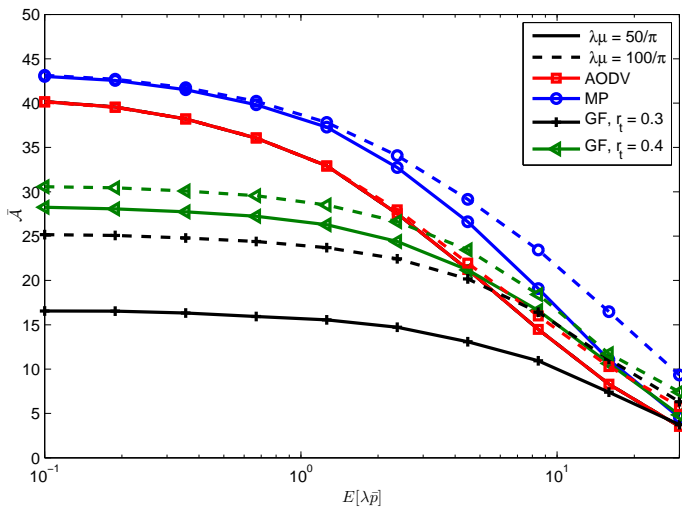}
\vspace{-0.5 cm}
\caption{Area spectral efficiency for each routing protocol when
$||X_{M\text{+1}}-X_{0}||=0.5$ as a function of the contention
density\emph{\ }with the relay density as a parameter. \label{Fig13_ASE_lambda}}
\end{figure}

Fig. \ref{Fig12_Reliability_lambda} and Fig. \ref{Fig13_ASE_lambda} illustrate the average path reliability and the area spectral
efficiency, respectively, for each routing protocol when $||X_{M\text{+1}%
}-X_{0}||=0.5$ as a function of the contention density $\mathbb E [ \lambda \overline{p} ]$
with the relay density $\lambda \mu$ as a parameter. The figures indicate the
degree to which an increase in the contention density is mitigated by an
increase in the relay density. The figures were generated by varying $p$ and
$\mu$ while maintaining $M=200$ and $\lambda=201/\pi.$ However, nearly the
same plots are obtained by varying $M$ or $\lambda$ while maintaining $p=0.3$
and $\mu=0.4.$ Thus, the contention density and relay density are of primary
importance, not the individual factors $p$, $\mu$, and $\lambda$.%


The figures can be used to determine which protocols are suitable for
achieving the performance requirements. For example, if the average path
reliability is required to be 0.9 when $||X_{M\text{+1}}-X_{0}||=0.5$, then
Fig. \ref{Fig12_Reliability_lambda} indicates that the maximum progress protocol or greedy forwarding
protocol is necessary. If in addition the relay density is $\lambda  \mu
=50/\pi,$ then the maximum progress protocol and the greedy forwarding
protocol with $r_{t}=0.4$ meet the requirement, and $\mathbb E [\lambda \bar p] \leq 1$ can be accommodated.

For both values of the relay density, the AODV protocol exhibits a lower
average path reliability and area spectral efficiency than the maximum
progress protocol. The greedy forwarding protocol with $r_{t}=0.3$ has a poor
average path reliability and area spectral efficiency when the relay density
is low because of the paucity of potential relays within its transmission
range. When $r_{t}=0.4$ and both the relay density and contention density are
large, the greedy forwarding provides an area spectral efficiency superior to
that of the AODV protocol and approaching that of the maximum progress protocol.

\section{Conclusions} \label{Sec:Conclusions}

This paper presents performance evaluations and comparisons of two geographic
routing protocols and the popular AODV protocol. The general methodology of
this paper can be used to provide a significantly improved analysis of
multihop routing protocols in ad hoc networks. Many unrealistic and improbable
assumptions and restrictions of existing analyses can be discarded.

The tradeoffs among the average path reliabilities, average conditional
delays, average conditional numbers of hops, and area spectral efficiencies
and the effects of various parameters have been shown for a typical ad hoc
network. Since acknowledgements are often lost due to the nonreciprocal
interference on the reverse paths, the AODV protocol has a relatively low path
reliability, and its implementation is costly because it requires a flooding
process. In terms of the examined performance measures, the greedy forwarding
protocol is advantageous when the separation between the source and
destination is small and the spreading factor is large, provided that the
transmission range and the relay density are adequate. The maximum progress
protocol is more resilient when the relay density is low and is advantageous
when the separation between the source and destination is large.

The approach presented in this paper can be readily extended to cover more
sophisticated network models. For instance, by incorporating a discrete-time
mobility model, systems with high mobility and carry-store-and-forward
networks can be accommodated. While the focus has been on unicast networks,
the extension to multicast is straightforward.

\vspace{-2.00 cm}

\balance
\vfill

\ifpdf
  \begin{IEEEbiography}{Don Torrieri}
\else
  \begin{IEEEbiography}[{\includegraphics[width=1in,height=1.25in,clip,keepaspectratio]{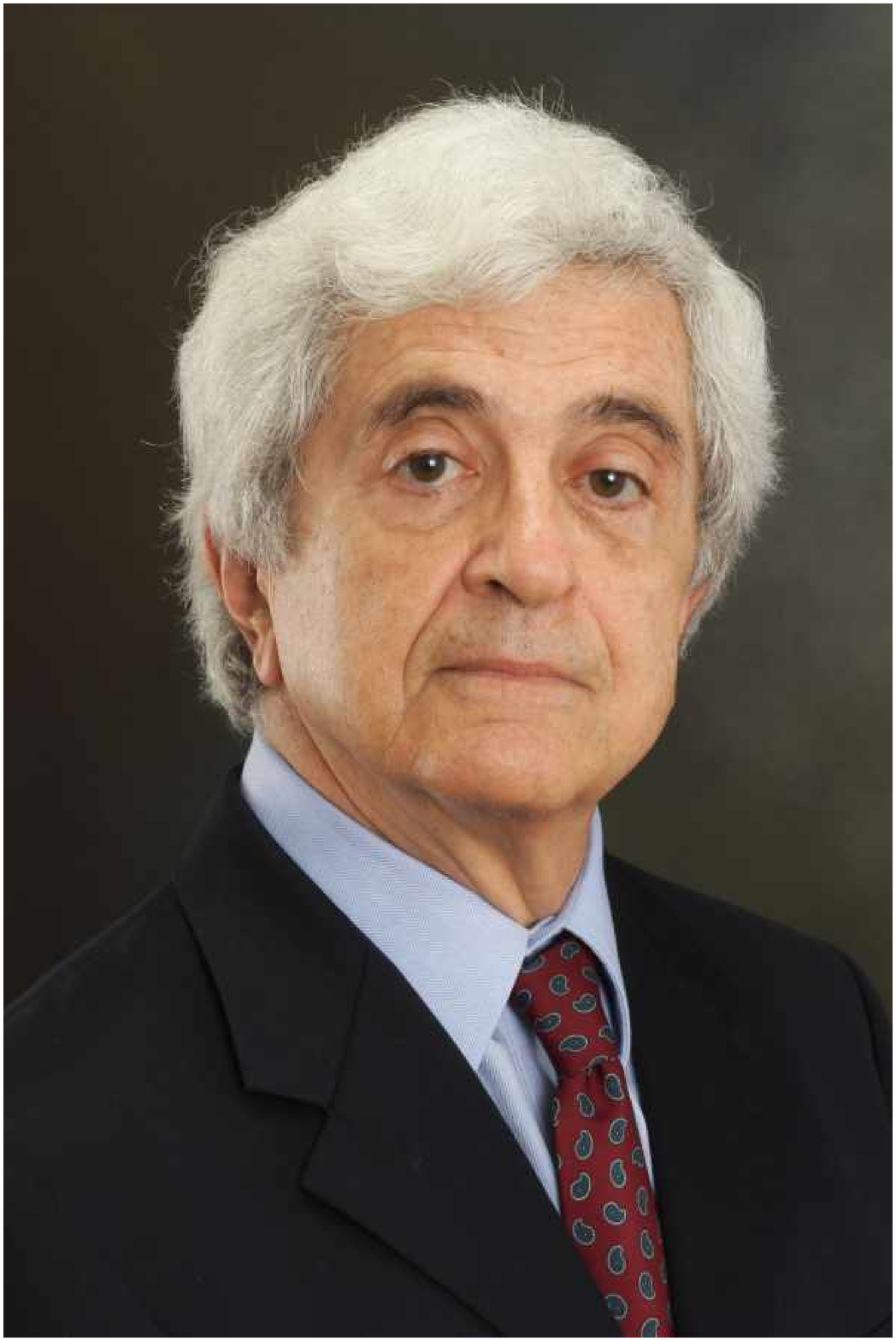}}]{Don Torrieri}
\fi
is a research engineer and Fellow of the US Army Research Laboratory. His primary research interests are communication systems, adaptive arrays, and signal processing. He received the B. S. degree from the Massachusetts
Institute of Technology and the M. S. and Ph.D. degrees from the University
of Maryland.  He is the author of many articles and several books including {\em Principles of Spread-Spectrum Communication Systems}, third ed. (Springer, 2015). He has taught many graduate courses at Johns Hopkins University and many short courses. In 2004, he received the IEEE Military Communications Conference Achievement award for sustained contributions to the field.
 In 2014, he received the Army Research Laboratory Publication Award.
\end{IEEEbiography}

\vspace{-1.00 cm}
\ifpdf
  \begin{IEEEbiography}{Salvatore Talarico}
\else
  \begin{IEEEbiography}[{\includegraphics[width=1in,height=1.25in,clip,keepaspectratio]{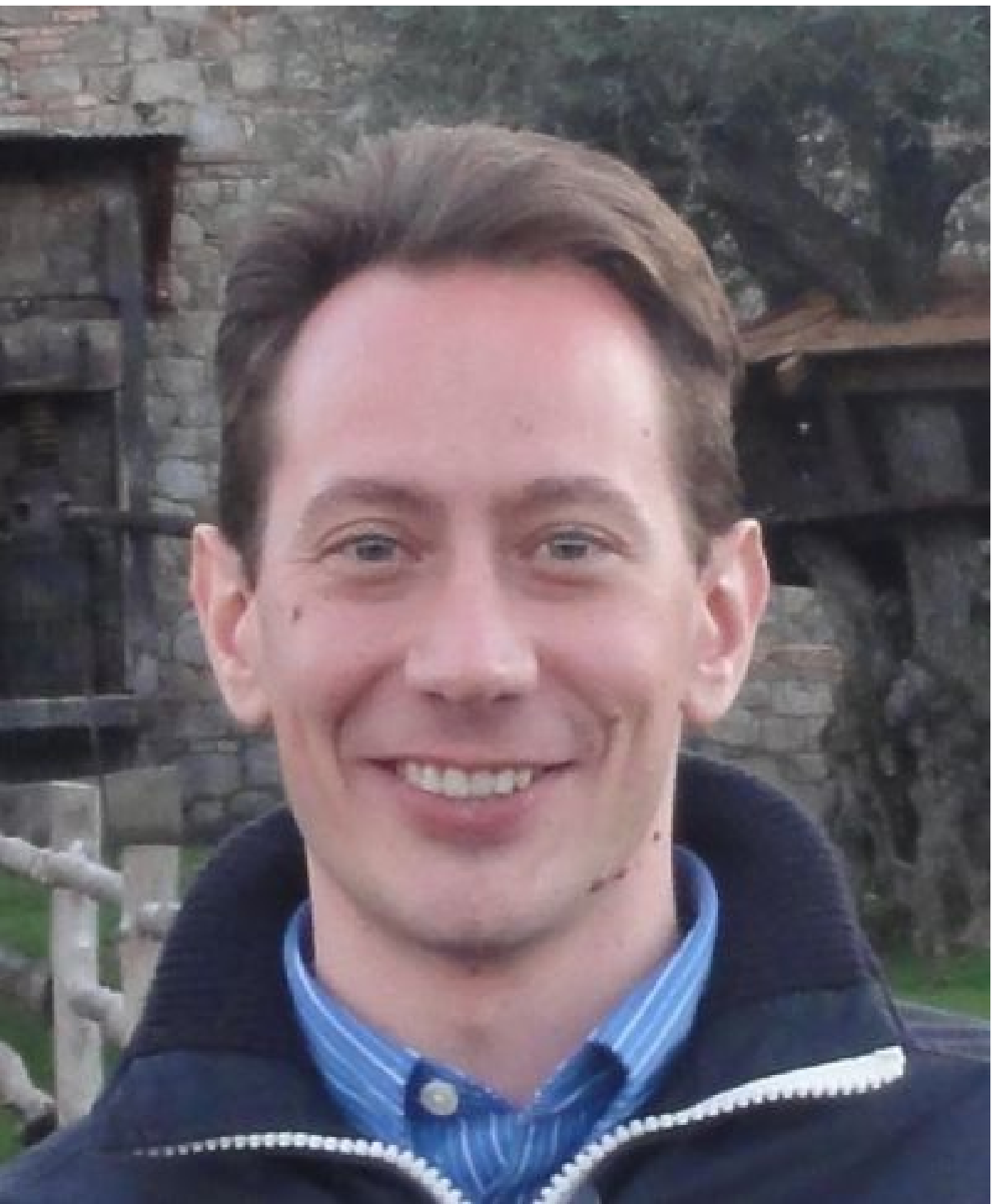}}]{Salvatore Talarico}
\fi
received the B.Sc and M.Sc. degrees in electrical engineering from University of Pisa, Pisa, Italy in 2006 and 2007 respectively, and the Ph.D in electrical engineering from West Virginia University, Morgantown, WV in 2015. From 2008 until 2010, before joining West Virginia University, he worked in the R$\&$D department of Screen Service Broadcasting Technologies (SSBT) as an RF System Engineer. His research interests lies in the area of software defined radio, information theory, and wireless communications. 
\end{IEEEbiography}

\vspace{-1.00 cm}
\ifpdf
  \begin{IEEEbiography}{Matthew C. Valenti}
\else
  \begin{IEEEbiography}[{\includegraphics[width=1in,height=1.25in,clip,keepaspectratio]{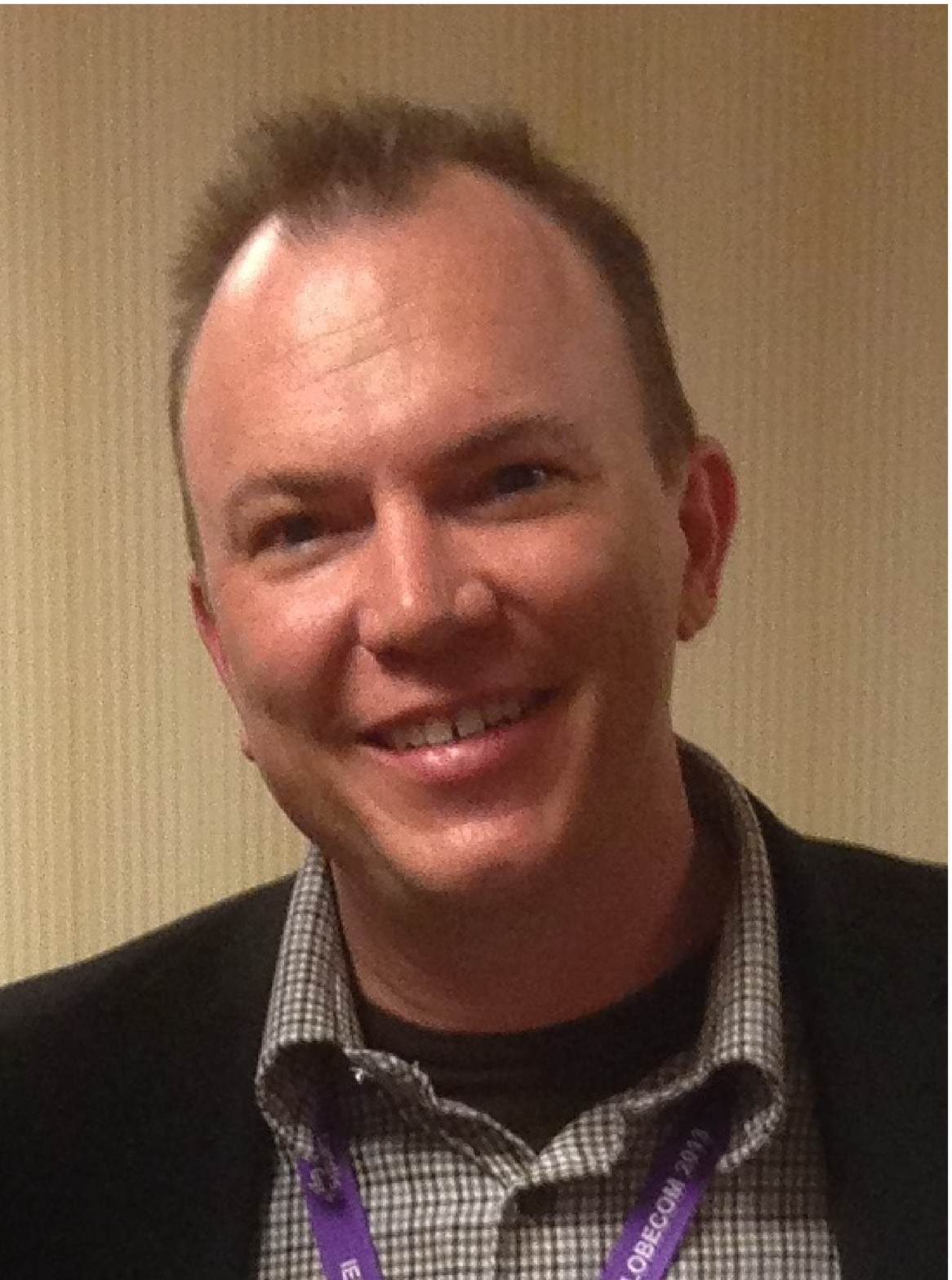}}]{Matthew C. Valenti}
\fi
is a Professor in the Lane Department of Computer Science and Electrical Engineering at West Virginia University and site director for the Center for Identification Technology Research (CITeR), an NSF Industry/University Cooperative Research Center (I/UCRC).  His research is in the area of wireless communications, including cellular networks, military communication systems, sensor networks, and coded modulation for satellite communications.   He is active in the organization of major conferences, including serving as the Technical Program Vice Chair for MILCOM 2015 and Globecom 2013, and as a track or symposium chair for MILCOM ('10,'12,'14), ICC ('09,'11), Globecom ('15), and VTC ('07).   He is an Executive Editor for IEEE Transactions on Wireless Communications, an Editor for IEEE Transactions on Communications, and the Chair of ComSoc's Communication Theory Technical Committee. Dr. Valenti is an alumnus of Virginia Tech, having received his BSEE in 1992 and Ph.D. in 1999, under the support of the Bradley Fellowship.  In addition, he received a MSEE from Johns Hopkins and worked as an Electronics Engineer at the US Naval Research Laboratory. He is registered as a Professional Engineer in the state of West Virginia.
\end{IEEEbiography}

\end{document}